\documentclass{mn2e}
\usepackage{psfig}
\usepackage{epsfig}

\newif\ifAMStwofonts

\def\sqiglt{\hbox{\rlap{\lower.55ex \hbox {$\sim$}}\kern-.05em \raise.4ex \hbox{$<$}\,}}
\def\sqiggt{\hbox{\rlap{\lower.55ex \hbox {$\sim$}}\kern-.05em \raise.4ex \hbox{$>$}\,}}

\def\til{\ensuremath{\sim\,}}

\def\chisq{\ensuremath{\chi^2}}
\def\rchisq{\ensuremath{\chi_{\nu}^{2}}}

\newcommand{\tim}[1]{\ensuremath{\times 10^{#1}}}
\def\deg{\ensuremath{^{\circ}}}
\def\etal{et al.\ }

\def\mekal{{\sc mekal}}

\title[The intermediate polar V405 Aur]{Why does the intermediate
polar V405 Aurigae show a double-peaked spin pulse?}

\author[Evans \& Hellier]{P.A. Evans\thanks{pae@astro.keele.ac.uk}
and Coel Hellier\\ Astrophysics Group, School of Chemistry and
Physics, Keele University, Staffordshire, ST5 5BG}

\date{Accepted 
      Received }

\pagerange{\pageref{firstpage}--\pageref{lastpage}}
\pubyear{2004}

\begin{document}

\maketitle

\label{firstpage}

\begin{abstract} 
V405~Aurigae is an intermediate polar showing a double-peaked
pulsation in soft X-rays and a single-peaked pulsation in harder
X-rays. From \emph{XMM-Newton\/} observations we find that the soft
band is dominated by blackbody emission from the heated white-dwarf
surface. Such emission is at a maximum when either magnetic pole
points towards us, explaining the double-peaked pulsation. The
symmetry of the pulses requires that the angle between the magnetic
and spin axes be high.

The single-peaked pulsation in harder X-rays is explained in the
usual way, as a result of opacity in the accretion curtains. 
However, the high dipole inclination means that the accretion
curtains are nearly in the plane. Thus the outer regions of the
curtains do not cross the line of sight to the accretion footprints,
explaining the absence of the deep absorption dip characteristic of
many intermediate polars. The sawtooth profile of this pulsation
requires that the magnetic axis be offset from the white-dwarf
centre.

We remark also on the double-peaked optical emission in this star. 
We suggest that the difference between V405~Aur's spin pulse and
those of other intermediate polars is the result of its high dipole
inclination.
\end{abstract}

\begin{keywords}
accretion, accretion discs -- stars: individual: V405~Aur
(RX\,J0558.0+5353) -- novae, cataclysmic variables -- X-rays:
binaries.
\end{keywords}

\section{Introduction}
\label{sec:intro}

V405 Aurigae (RX\,J0558.0+5353) was discovered in the \emph{Rosat}
All-Sky Survey and identified as an intermediate polar (a cataclysmic 
variable with a magnetic white-dwarf primary) by Haberl \etal (1994).

It is notable, firstly, for showing a soft blackbody component in
the X-ray spectrum, one of a number of such objects discovered with
\emph{Rosat}. Secondly, its soft-X-ray and optical emission shows a
double-peaked modulation at the white-dwarf spin period (e.g.\ Allan
\etal1996), whereas most of these stars show a
single-peaked modulation (see, e.g., Patterson 1994 or Hellier 2001
for reviews of this class).

The hard X-ray emission in intermediate polars (IPs) originates below
a stand-off accretion shock near the magnetic poles of the white
dwarf. The soft blackbody emission is then understood as arising
from heated white-dwarf surface around the accretion footprints. 
This is nearly always seen in the AM~Her class of cataclysmic
variable, but it is seen only in some IPs, for which the reason 
is unclear. 

The issue of why some IPs show a single-peaked pulsation, whereas
others show a double-peaked pulsation, is also unclear. One idea
(e.g.\ Hellier 1996; Allan \etal1996; Norton \etal 1999) notes that
IPs with shorter spin periods will have smaller magnetospheres in
which the accretion discs are disrupted nearer the white dwarf. This
could result in shorter, fatter `accretion curtains' of material
which might have lower opacity in the vertical direction, thus
preferentially beaming X-rays along magnetic field lines. The two
magnetic poles would combine to produce a double-peaked pulsation. 
With longer spin periods, where disc disruption occurs further out,
the opposite might hold, with tall, thin accretion curtains 
preferentially beaming X-rays out of the sides. The two poles would 
then act in phase, producing a single-peaked pulsation. 

The \emph{XMM-Newton\/} X-ray satellite has a larger collecting area
and better spectral resolution than \emph{Rosat\/}, allowing us to
return to V405~Aur with better X-ray data than previously obtained.
We report here on a 30-ks \emph{XMM-Newton\/} observation aimed at
understanding the pulsation at the 545-s spin period of V405~Aur.

\section{Observations and lightcurves}
\label{sec:obsphot}

V405~Aur was observed by the \emph{XMM-Newton\/} satellite (Jansen
\etal2001; Turner \etal2001) on 2001 October 5. A 30-ks observation was 
made with the EPIC-MOS cameras in {\sc timing uncompressed mode} 
using the Thin Filter 1; the PN camera was not in operation.
The source was outside the OM window, so these data were not
used. 

We analysed the data using the {\sc xmm-sas} software v5.4.1.
In {\sc timing mode} the central CCD data is compressed onto a single
axis, thus selection is based on columns rather than pixels. Only
single or double pixel events with a zero quality flag were selected
from a 44-column region centred on the source. We also removed column
315 from the MOS-2 data, which showed a spurious signal at low
energies. Owing to the limited coverage of the {\sc timing mode} 
window we used an adjacent chip (CCD3) to estimate the background. 

\begin{figure}
\begin{center}
\psfig{file=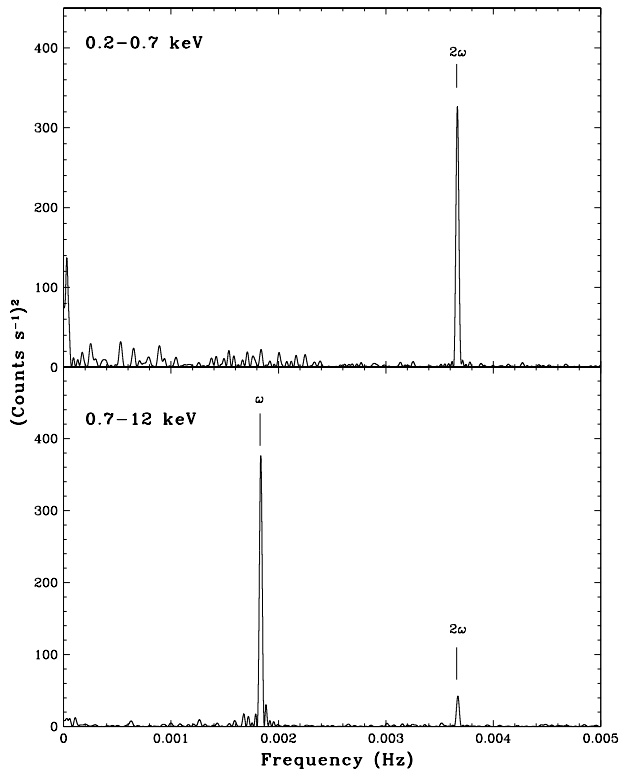,width=8.1cm}
\caption{Power spectra of the X-ray data (MOS-1 + MOS-2) in the soft
(upper panel) and hard (lower panel) energy bands.}
\label{fig:ft}
\end{center}
\end{figure}

The power spectrum of the 0.2--12 keV data from the MOS cameras is
given in Fig.~\ref{fig:ft}. This shows prominent peaks at the spin
frequency and its first harmonic (periods 545.5 s and 272.7 s
respectively). We find that the first harmonic is the stronger at
energies below 0.7 keV (hereafter the `soft' band), while the
fundamental is the stronger at energies above 0.7 keV (hereafter the
`hard' band), as previously reported by Allan \etal(1996).

\begin{figure}
\begin{center}
\psfig{file=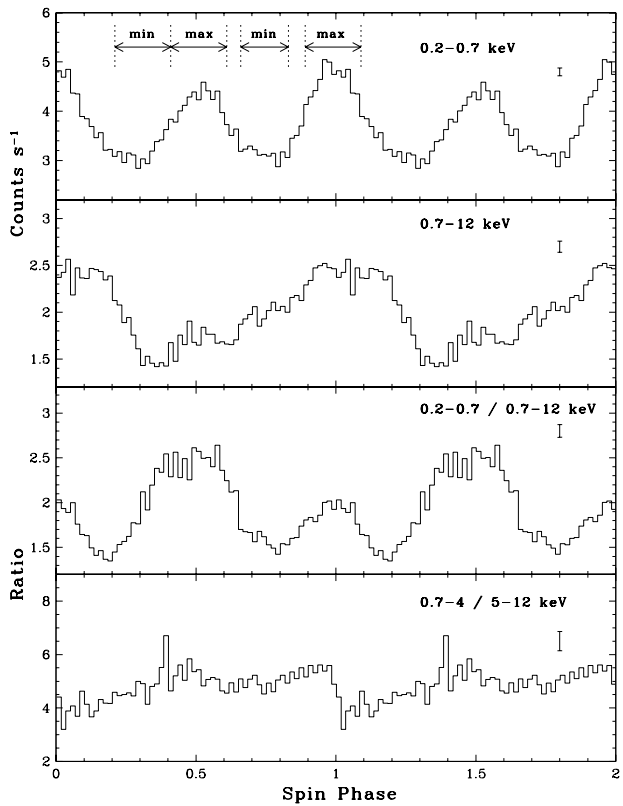,width=8.1cm}
\caption{The X-ray data (MOS-1 + MOS-2) from the soft and hard bands
(top panels), folded on the 545-s spin cycle. The lower panels contain the
0.2--0.7\,/\,0.7--12 keV and 0.7--4\,/\,5--12 keV softness ratios.
Typical error bars are shown. Phase zero is taken as the peak of the
larger of the soft maxima, and corresponds to HJD~2452187.55904.}
\label{fig:spin}
\end{center}
\end{figure}

The folded lightcurves from the two bands (Fig.~\ref{fig:spin})
confirm these results, showing that the soft emission has a
double-peaked profile whereas the hard emission has a
single-peaked, sawtooth modulation. 

\section{Spectroscopy}
\label{sec:spec}

\begin{figure}
\begin{center}
\psfig{file=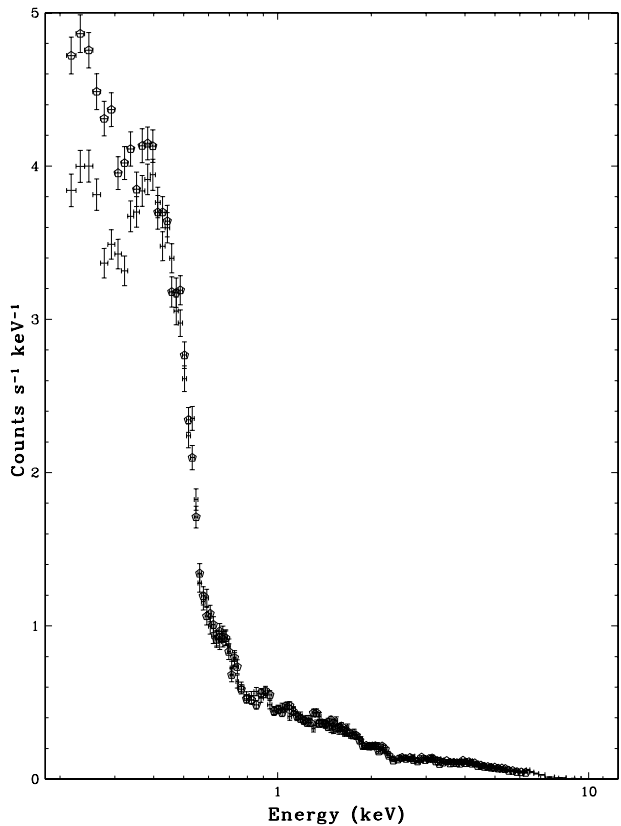,width=8.1cm}
\caption{The phase-averaged spectra from the MOS-1 and MOS-2 (circles)
cameras. The systematic discrepancy below 0.4 keV is clear.}
\label{fig:system}
\end{center}
\end{figure}

Little {\sc timing mode} data has yet been published, and we used
pre-release versions of new quantum efficiency calibration files and
canned response matrices (Sembay, private communication). However,
there were clear calibration discrepancies between the two EPIC-MOS
instruments below 0.4 keV (Fig.~\ref{fig:system}). We thus added
systematic errors of 8\%\ to the data in this energy range. Other
than this, we do not make allowance for calibration uncertainties in
calculating \chisq\ values. The data from the two cameras were fitted
simultaneously in the fits reported below.

The X-ray emission in an IP arises from plasma heated to X-ray
temperatures at a stand-off accretion shock, which then cools as it
approaches the white dwarf surface (e.g.\ Aizu 1973; Cropper
\etal1999). The {\sc cemekal} model reproduces a multi-temperature
spectrum, with a power-law distribution of temperatures, based on the
{\sc mekal} plasma model. 

Fitting our phase-averaged spectrum with a {\sc cemekal} model plus
simple and partial-covering absorption (as commonly found in IPs)
gave a poor fit (\rchisq=20) owing to a large soft excess. Adding a
blackbody emitter to model the soft emission improved the fit hugely
to \rchisq=1.38.

We then tried replacing the {\sc cemekal} with two \mekal\ components
and found a significant improvement in fit quality (\chisq=1276,
\rchisq=1.28) implying that a power-law is a poor description of the
temperature distribution in the plasma column. The best-fitting
temperatures were 0.2 and 9 keV. The blackbody component was still
necessary (\rchisq=1.60 without it), as were both simple and
partial-covering absorption (\rchisq=1.71 without these).  There were
significant residuals at the 6.4-keV iron fluorescence line, so we
added a narrow Gaussian component, reducing \chisq\ to 1203
(\rchisq=1.21).

In the above models the simple and partial-covering absorbers acted
on all of the emission components. However, the partial-covering
absorber is sufficiently dense as to completely block any blackbody
emission that it covers, thus its effect is redundant with a change
in blackbody normalisation. So as a final tweak we altered the model
so that the blackbody was only absorbed by a simple absorber, which,
for generality, was allowed to have a different column than the
absorber acting on the \mekal s. This yielded a slight \chisq\
improvement (\chisq=1187, \rchisq=1.19).

\begin{figure*}
\begin{center}
\psfig{file=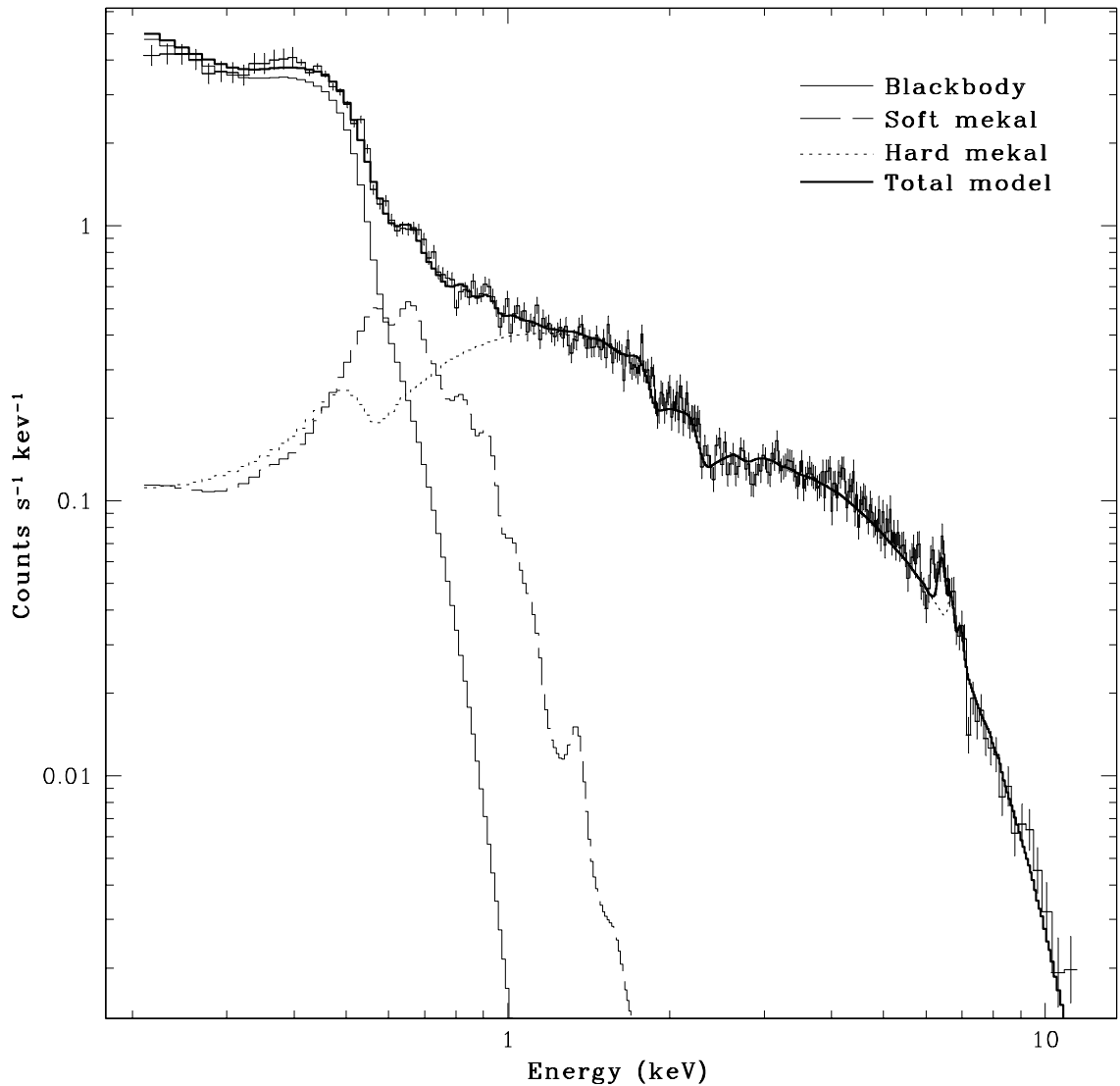,width=15cm}
\caption{The phase-averaged spectrum with the fitted model, and
the contributions from each component.}
\label{fig:comps}
\end{center}
\end{figure*}

The resultant model is given in Table~\ref{tab:average} and
illustrated in Fig.~\ref{fig:comps}. Note that the spectral
parameters vary over spin phase, so those listed in
Table~\ref{tab:average} will be weighted averages.

\begin{table}
\begin{center}
\begin{tabular}{lccc}
\hline
Component &  Parameter          & Value          & Error \\ 
          &  (Units) \\
\hline
Absn.     &  $n_{\rm H}$ (cm$^{-2}$)  & 1.06\tim{21}      & (+0.09, --0.12) \\ 
Blackbody &  $kT$ (keV)         & 3.97\tim{-2}   & (+0.38, --0.42)\\ 
          &  Normalisation      & 1.34\tim{-2}   & (+1.64,--0.79)\\ 
Absn      &  $n_{\rm H}$ (cm$^{-2}$)  & 2.06\tim{21}      & (+0.29, --0.29) \\
Part. Absn.  &  $n_{\rm H}$ (cm$^{-2}$)  & 6.08\tim{22}   & (+0.58, --0.52) \\
             &  CvrFract        & 0.52           & (+0.03, --0.03) \\
Gaussian  &  Energy (keV)       & 6.42           & (+0.02, --0.02) \\
          &  Normalisation      & 1.96\tim{-5}   & (+0.44, --0.45) \\
          &  Eq. Width (eV)     & 121            & (+27, --28) \\
Mekal     &  $kT$ (keV)         & 0.168          & (+0.008, --0.003) \\
          &  Abundance          & 3.39\tim{-2}   & (+1.14, --0.62) \\
          &  Normalisation      & 0.32           & (+0.16, --0.12) \\
Mekal     &  $kT$ (keV)         & 9.00           & (+0.92, --0.73) \\
          &  Abundance          & 0.12           & (+0.35, --0.36) \\
          &  Normalisation      & 1.31\tim{-2}   & (+0.08, --0.08) \\
\hline
\end{tabular}
\caption{Model components and parameters fitted to the phase-averaged
spectrum (see Section~\ref{sec:spec}). The errors are the 90\%
confidence errors according to formal statistics (to the same power
of ten as the values); note however that these are likely
underestimates, since they do not account for the calibration
systematics.}
\label{tab:average}
\end{center}
\end{table}

\section{Phase-resolved spectroscopy}
\label{sec:resspec}

We have found that below 0.7 keV the spectrum is dominated by a
blackbody component, whereas above 0.7 keV the spectrum is dominated
by optically thin plasma (Section~\ref{sec:spec}). Further,  the soft
blackbody component is double-peaked over the spin cycle, whereas the
higher-temperature plasma has a single-peaked, sawtooth pulse profile
(Fig.~\ref{fig:spin}). Thus, for phase-resolved spectroscopy, it
makes sense to deal with the soft pulse and the hard pulse
separately. 

\subsection{The soft pulse}
\label{sec:softspec}

To analyse the soft pulse we defined phase regions encompassing the
maxima (phases 0.41--0.61 and 0.89--1.09) and minima (phases
0.21--0.41 and 0.66--0.83), as illustrated in Fig.~\ref{fig:spin}.

We fitted all four phase regions simultaneously with the model
developed above (Section~\ref{sec:spec}), which comprised a soft
blackbody and two \mekal\ emitters, with a simple absorber acting on
the blackbody, and both simple and partial-covering absorbers acting
on the \mekal\ emitters. For analysing the soft pulse the fits were
allowed to optimise for the entire spectrum, but the quoted \chisq\
values were calculated only from 0.2--0.7 keV (the changes in \chisq\
between models were similar if \chisq\ was calculated over the entire
energy range).

Allowing the model to optimise for each phase region separately
yielded a best-fitting \chisq\ of 263 (\rchisq=1.27). Constraining
the simple absorber to be the same at all phases caused minimal
change in fit quality but reduces the number of free parameters
(\chisq=264, \rchisq=1.25).

Alternatively, allowing the absorption to vary freely, but forcing
the blackbody emitter to remain constant between phase regions, 
yielded a \chisq\ of 308.51 (\rchisq=1.44). This increase suggests
that the soft modulation is a change in blackbody normalisation
rather than a change in absorption. 

Note that the 0.2--0.7 keV region is the worst affected by
calibration uncertainties, so the quoted \chisq\ values are
unreliable. However, if we fit the data from each camera separately,
without adding systematic uncertainties, we reach the same
conclusion: that the soft pulse is mainly a change in visible
emitting area, and not a change in absorption. 

\subsection{The Hard Pulse}
\label{sec:hardspec}

Since the 0.2--0.7\,/\,0.7--12 keV softness ratio is effectively a
ratio of the blackbody emission to the plasma emission, we have also
computed the 0.7--4\,/\,5--12 keV softness ratio, which is sensitive
to spectral changes in the harder plasma emission. 

This ratio (Fig.~\ref{fig:spin}) shows little variation other than a
hardening at phases \til0.0--0.2, near the maximum of the hard pulse.
We thus extracted and compared spectra for the phase regions 0.0--0.2
and 0.4--0.9. 

Fitting the model to the two phase regions independently gave a total
\chisq\ of 1253 (\rchisq=1.10; in this section \chisq\ values are
for  energies above 0.7 keV only; again, this does not affect the
$\Delta\chisq$ values). Constraining the absorption (simple and
partial covering) to have the same value in each phase region gave an
identical \chisq\ of 1253, implying no absorption change.
Alternatively, requiring the normalisations of the \mekal s to be the
same in each region worsened \chisq\ to 1530 (\rchisq = 1.34).

Thus we conclude that the hardening near flux maximum is caused by a
change in the ratio of the soft and hard \mekal\ components. At all
other phase regions there appears to be little or no change in the
ratios of the two \mekal\ components, nor in the absorption. This is
despite the fact that the overall flux level changes considerably
over these phase regions. Similar results, that the hard pulse does
not show much energy dependence, were reported using {\it BeppoSAX\/}
data by de Martino \etal(2004).

\section{Discussion}

\begin{figure*}
\begin{center}
\psfig{file=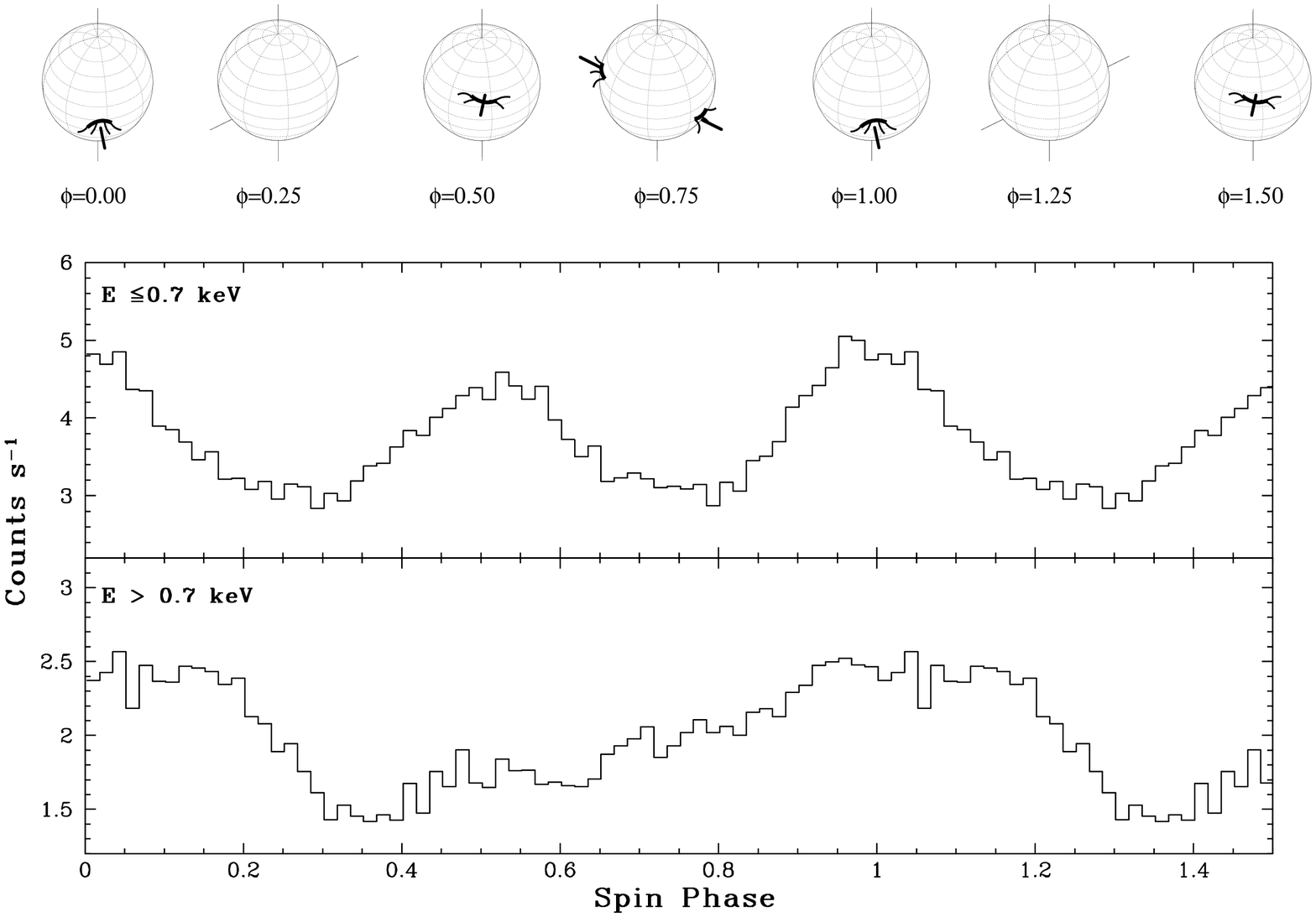,width=16cm}
\caption{Schematic diagram showing the location of the emitting
regions and magnetic poles at various phases. The dipole offset has
been exaggerated. The arc-shaped regions are the blackbody emitting
areas and the short lines depict the accreting field lines.}
\label{fig:schem}
\end{center}
\end{figure*}

V405~Aurigae has two main spectral components in the X-ray band. 
Below 0.7 keV the spectrum is dominated by a blackbody component 
which shows a double-peaked modulation over the spin cycle. Above 0.7
keV it is dominated by thermal plasma emission which has a
single-peaked, sawtooth modulation over the spin cycle.

IP X-ray modulations are often the result of absorption as `accretion
curtains' of material sweep across the line of sight with the spin
cycle (e.g.\ Hellier, Cropper \&\ Mason 1991; Kim \&\ Beuermann
1995). Further, it has been proposed that whether the pulsation is
single-peaked or double-peaked depends on whether the accretion
column is short and fat, with greater horizontal opacity, or tall and
thin, with greater vertical opacity  (Hellier 1996; Allan \etal1996;
Norton \etal1999).

V405~Aur, however, does not fit the above model since, firstly, it
shows both single-peaked and double-peaked behavior in the same star.
Secondly, our spectral modelling shows that the absorption varies
little over spin phase (Section~\ref{sec:spec}), which is the essence
of the above model. 

\subsection{Explaining the spin pulse}

We suggest that the modulation of the blackbody emission is simply
the foreshortening of the accreting polecaps, which are viewed face
on when a pole points towards us (maximum) and foreshortened when the
pole is near the white dwarf limb (minimum). Thus one of the poles
is  in the middle of the visible face at each maximum (see Fig.~5).
The difference between the two maxima is slight, so we require that
both accretion polecaps be near the equator of the white dwarf,  so
that there is little difference in their visibility. This implies
that the angle between the magnetic and spin axes is large, as is 
also suggested by the polarimetry reported by Shakhovskoj, Andronov
\&\ Kolesnikov (2001). 

Turning now to the hard X-ray pulse, the accretion geometry for a
highly inclined dipole is illustrated in Fig.~\ref{fig:schem2}. This
shows that the difference in view between the accretion footprints at
the upper and lower poles is relatively small. Nevertheless, we can
adopt the standard accretion-curtain model (see, e.g., Hellier
\etal1991) and suppose that the lower pole viewed at phase 0 is
brighter than the upper pole viewed at phase 0.5
(Fig.~\ref{fig:schem2}). This is because the opacity in the column
causes X-rays to emerge preferentially perpendicular to the curtain,
which means that the lower pole is somewhat better presented. 

Note that with such a high dipole inclination, the cooler, outer
regions of the accretion curtains never cross the line of sight. This
explains why we don't see a prominent absorption dip dominating the
spin pulse, in contrast to many other IPs (e.g.\ AO~Psc, Hellier
\etal1991). However, the X-rays will be emerging from a highly
ionized post-shock region in which absorption and electron scattering
will cause opacity, thus explaining the change in brightness with
aspect of the accretion curtain, and the need for a partial-covering
absorber in the spectral fit.

Note that at phase 0 the blackbody emission seen at the lower pole
would be from heated surface on the side of the curtain further from
the magnetic pole (that nearer the pole being obscured by the
curtain) whereas at phase 0.5 the blackbody emission at the upper
pole is from the side nearer to the magnetic pole (see
Fig.~\ref{fig:schem2}).   Thus an asymmetry between these regions can
explain the fact that one maximum is higher than the other. 

\subsection{The sawtooth profile}

We further need to explain the fact that the hard-X-ray pulse shows a
sawtooth profile, declining more rapidly than it rises. We suggest
that this can be explained if the magnetic dipole is offset by \til
0.1--0.2 white-dwarf radii. Such offsets are commonly found in
magnetic white dwarfs (Wickramasinghe \&\ Ferrario 2000). 

From the deductions above, the lower pole is on the meridian facing
us at phase 1 (thus presenting maximum blackbody area and producing a
soft maximum). However, suppose that the dipole is offset to the left
(Fig.~\ref{fig:schem}), so that the magnetic pole only points towards
us shortly afterwards. This will slightly delay the hard X-ray peak. 

The lower pole then disappears over the limb at phase 0.25. Because
of the offset, the upper pole does not yet appear, and this 
asymmetry explains the relatively rapid fall in flux at this phase. 

With the asymmetry pushing both poles to the far side of the white
dwarf, we have lowest flux occurring before 0.5. Then the upper pole
appears and is in the middle of the face near phase 0.5.  Its less
favorable aspect makes the system fainter than when we see the lower
pole at phase 1.0. 

As we continue through the cycle towards phase 0.75, the offset of
the dipole biases both poles towards the visible face of the white
dwarf, which is why we see more flux approaching phase 0.75 than we
do following phase 0.25, thus explaining the sawtooth. However, since
both accretion curtains are then presented with their azimuthal sweep
being edge on to us, the flux level is lower than at phase 1.0.

\subsection{The optical spin pulse}
\label{sec:opt}

Although we do not present optical data in this paper, we can 
consider whether the model can explain the double-peaked nature of
the optical spin pulse (e.g.\ Allan \etal1996;  Skillman 1996; Still,
Duck \&\ Marsh 1998). 

In the standard accretion-curtain model, with a relatively low dipole
inclination, the changing visibility of the curtains results in a
single-peaked pulse with a maximum when the upper pole points away
from us (Hellier \etal1991; Kim \&\ Beuerman 1996). 

\begin{figure}
\begin{center}
\psfig{file=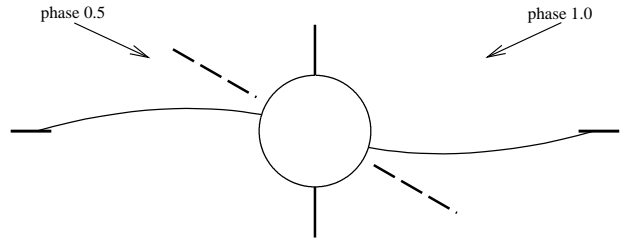,width=8.1cm}
\caption{Side-on schematic view of the model proposed in the text,
shown here for a system inclination of 65\deg\ and a dipole
inclination of 60\deg.}
\label{fig:schem2}
\end{center}
\end{figure}

However, in V405~Aur we are postulating a large dipole inclination,
such that the accretion curtains are nearly flat in the orbital plane
(Fig.~\ref{fig:schem2}). Note that this is also supported by the
optical polarimetry of Shakhovskoj \etal(2001) which implies equal
visibility of two poles near the equator. 

With accretion curtains nearly in the plane, the view when the upper
pole points towards us is similar to that when the lower pole points
towards us (Fig.~\ref{fig:schem2}). Thus we can explain a double-peaked
optical pulse if the curtains are brighter when one of the poles
points towards us and fainter when both are at quadrature (or vice
versa).

\subsection{The blackbody emitting area}
 
From our spectral fits we can estimate the blackbody emitting area.
Note, however, that the calculated blackbody flux is strongly
dependent upon the degree of absorption acting on it. In our best
fitting model (Table~\ref{tab:average}) this is surprisingly large
(\til10$^{21}$ cm$^{-2}$), particularly if we associate the simple
absorber with interstellar absorption. This in turn causes the flux
measured to be high. However, forcing the column density to take on a
lower value (e.g.\ 3\tim{20} cm$^{-2}$) causes a worsening in \chisq
of \sqiggt100. Note, though, that the energy range over which the
absorbed blackbody acts is that for which the calibration
uncertainties are at their worst, making the values  unreliable. In
what follows we use the flux calculated from the best fit ($n_{\rm H}
= 10^{21}$ cm$^{-2}$).

The blackbody emission component (Section~\ref{sec:spec}) has an
unabsorbed bolometric flux of 1.12\tim{-12} J m$^{-2}$ s$^{-1}$ with
a temperature of 40 eV. This implies an emitting area of 4.76\tim{5}
($d$/300 pc)$^{2}$ km$^{2}$, which corresponds to \til8\tim{-4} of a
white-dwarf surface [Haberl \&\ Motch (1995) reported a corresponding
value of 2\tim{-5}, based on a temperature of 57 eV and a column of
5.7\tim{20} cm$^{-2}$.] This area is consistent with other estimates
for the accretion area in IPs, for example the upper limit of 0.002
deduced by  Hellier (1997) from eclipse timings of XY~Ari.

\section{Conclusions}
\label{sec:conc}

\emph{XMM-Newton\/} observations of V405~Aur confirms that the soft
X-rays are dominated by a blackbody component, presumably from heated
white-dwarf surface surrounding the accretion regions. The heated
region covers \til8\tim{-4} of the white dwarf, comparable to the
upper limit found from eclipse timings of XY~Ari. 

We propose that the modulation of this component results from
foreshortening of the blackbody emitting regions. We view the heated
surface most favourably when either pole points towards us, so we see
a double-peaked modulation in soft X-rays. The equality of the two
maxima requires that the magnetic axis be highly inclined from the
spin axis.

The hard X-ray emission shows a single-peaked, sawtooth modulation.
This does not show the strong energy dependence of absorption, 
as usual in IPs. We suggest that this is because, with the high
dipole inclination, the outer parts of the accretion curtains
never cross the line of sight. However, electron scattering and
opacity in the highly ionized post-shock column causes the intensity
variation with spin phase.  The sawtooth shape of the pulsation requires 
that the magnetic axis be offset from the white-dwarf centre. 

We also suggest that the high dipole inclination is responsible for
the double-peaked optical pulse.  The high dipole inclination appears
to be the main reason for the differences between V405~Aur's
pulsation and typical IP behaviour.

\section*{Acknowledgements}

The observation reported here was performed by Leicester University,
who also constructed the MOS cameras and developed the analysis
software. In particular we thank Steve Sembay and Richard Saxton for
their help in overcoming calibration difficulties, and for providing
us with the latest pre-release CCF and RMF files for use in this
analysis. We also thank Liza van Zyl and Gavin Ramsay for some
illuminating discussions.

\label{lastpage}
\end{document}